\documentclass[lettersize,journal]{IEEEtran}
\usepackage{amsmath,amsfonts}
\usepackage{algorithmic}
\usepackage{algorithm}
\usepackage{array}
\usepackage[caption=false,font=normalsize,labelfont=sf,textfont=sf]{subfig}
\usepackage{textcomp}
\usepackage{stfloats}
\usepackage{url}
\usepackage{multirow}
\usepackage{tabularx,booktabs} 
\usepackage{verbatim}
\usepackage{graphicx}
\usepackage{cite}
\usepackage{makecell}
\usepackage{url}
\usepackage{relsize} 
\usepackage{color}
\makeatletter
\let\NAT@parse\undefined
\makeatother
\usepackage{hyperref}

\begin{document}

\title{RIS-aided Wireless-Powered Backscatter Communications for Sustainable Internet of Underground Things}

\author{Kaiqiang~Lin and Yijie Mao

\thanks{This work was supported in part by the National Nature Science
Foundation of China under Grant 62201347; and in part by Shanghai Sailing
Program under Grant 22YF1428400. (\textit{Corresponding author: Yijie Mao})}  

\thanks{K. Lin and Y. Mao are with the School of Information Science and Technology, ShanghaiTech University, Shanghai, China.}

\thanks{K. Lin is also with the Department of Computer, Electrical and Mathematical Sciences and Engineering (CEMSE), King Abdullah University of Science and Technology (KAUST), Jeddah, Kingdom of Saudi Arabia.}
}

\maketitle
\begin{abstract}
Wireless-powered underground sensor networks (WPUSNs), which enable wireless energy transfer to sensors located underground, is a promising approach for establishing sustainable internet of underground things (IoUT). To support urgent information transmission and improve resource utilization within WPUSNs, backscatter communication (BC) is introduced, resulting in what is known as wireless-powered backscatter underground sensor networks (WPBUSNs). Nevertheless, the performance of WPBUSNs is significantly limited by severe channel impairments caused by the underground soil and the blockage of direct links. To overcome this challenge, in this work, we propose integrating reconfigurable intelligent surface (RIS) with WPBUSNs, leading to the development of RIS-aided WPBUSNs. We begin by reviewing the recent advancements in BC-WPUSNs and RIS. Then, we propose a general architecture of RIS-aided WPBUSNs across various IoUT scenarios, and discuss its advantages and implementation challenges. To illustrate the effectiveness of RIS-aided WPBUSNs, we focus on a realistic farming case study, demonstrating that our proposed framework outperforms the three benchmarks in terms of the sum throughput. Finally, we discuss the open challenges and future research directions for translating this study into practical IoUT applications.
\end{abstract}

\section{Introduction}
\begin{figure}[!t]
\centering
\includegraphics[width=3.4in]{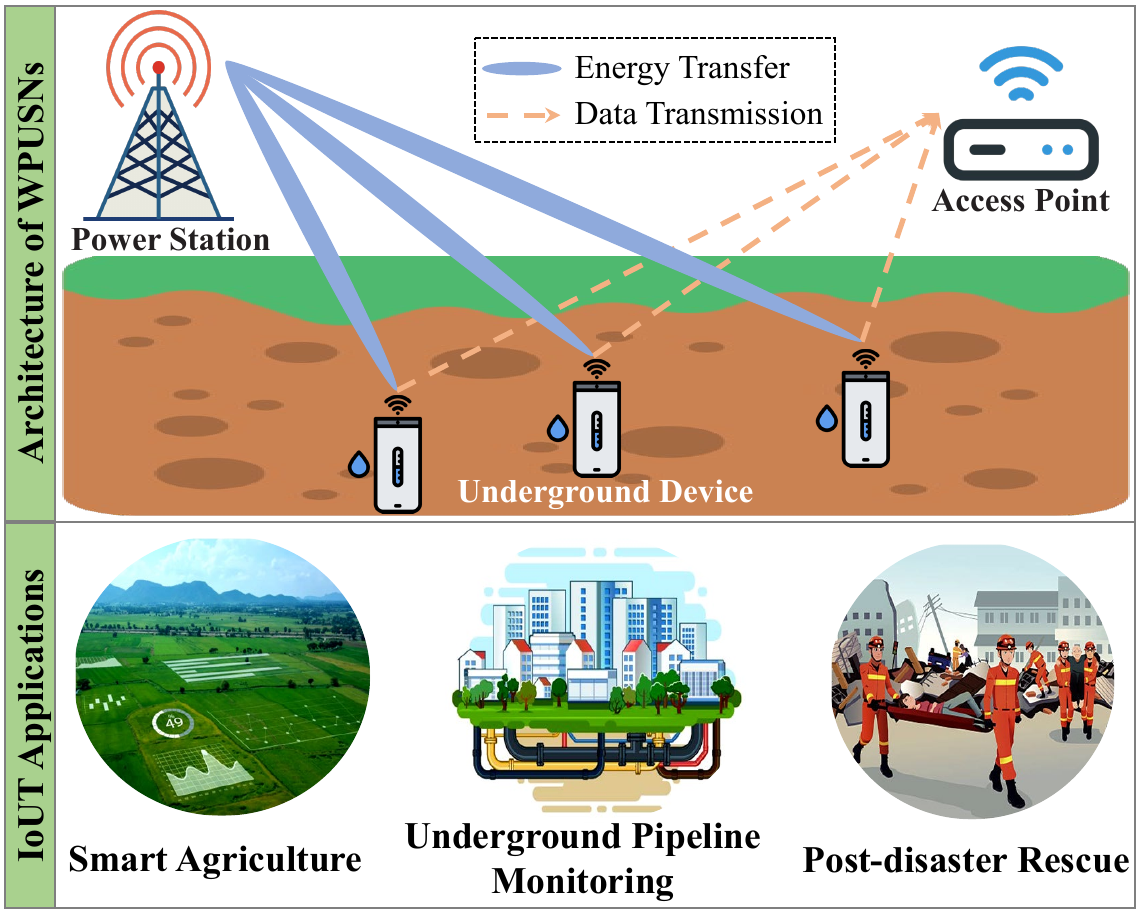}
\caption{A multi-user wireless-powered
underground sensor network (WPUSN) system for various IoUT applications.}
\label{fig_WPUSNs}
\end{figure}

\IEEEPARstart{I}{nternet} of Underground Things (IoUT) enables in-situ monitoring of underground resources via wirelessly connected underground devices (UDs), facilitating various applications such as smart agriculture, underground pipeline monitoring, and post-disaster rescue~\cite{LinWUSNsMag}. However, due to the high attenuation in underground soils, UDs require significantly more energy for reliable data transmission compared to terrestrial Internet of Things (IoT). With the limited battery capacity of these devices and the impracticality of regular battery replacement, achieving sustainable IoUT remains challenging. To address this, radio frequency (RF) wireless energy transfer (WET) technique has been implemented in the subterranean domain for recharging UDs, giving rise to wireless-powered underground sensor networks (WPUSNs)~\cite{LiuWPUSNs}, as illustrated in Fig.~\ref{fig_WPUSNs}. In a typical WPUSN system, UDs utilize the harvest-then-transmit (HTT) protocol, where they harvest energy from a power station (PS) before using the harvested energy to transmit the sensor data to an access point (AP) via wireless information transmission (WIT). As UDs requires a longer time to harvest energy for reliable WIT, the emergency information transmission cannot be easily guaranteed in WPUSNs, specially when the UDs are buried at high depths or located far from the PS. 

Backscatter communication (BC), which exploits the dedicated/ambient RF source to backscatter information, enables sustainable battery-free operations for IoT~\cite{BCMag}. The integration of BC with WPUSNs, as proposed in~\cite{LinBSWPUSNs}, introduces a new concept: BC-assisted WPUSNs, also known as wireless-powered backscatter underground sensor networks (WPBUSNs). This approach has been shown to facilitate the emergency information transmission and enhance the network throughput in WPUSNs. However, the performance benefits achieved through BC are usually degraded in high-attenuation scenarios, such as those involving increased burial depth required for underground monitoring applications, high volumetric water content (VWC) of soil following rainfall and irrigation, and a high probability of non-line-of-sight (NLOS) conditions. NLOS issues may arise from obstacles such as crops and vegetation in smart agriculture or trees and buildings in urban pipeline monitoring.

Recently, reconfigurable intelligent surface (RIS) has emerged as a promising technique for 6G. An RIS typically consists of a planar array of low-cost passive elements that intelligently manipulate the phase shifts of incoming signals, steering them in the desired direction to enhance received signal power and mitigate multiuser interference in terrestrial IoT applications~\cite{RISMag}. Therefore, RIS holds great potential to overcome the aforementioned challenges in WPBUSNs. However, while extensive research has delved into the benefits of RIS in terrestrial IoT, its performance gains remain unclear when a subterranean channel component is introduced. 

Motivated by this research gap, we explore the feasibility of synergizing RIS and WPBUSNs, leading to the development of RIS-aided WPBUSNs. To the best of the authors' knowledge, this work is the first one to assess the effectiveness of RIS in improving the sum throughput of WPBUSNs. Our main contributions are summarized as follows:
\begin{enumerate}
    \item We conceptualize a general architecture of RIS-aided WPBUSNs under different IoUT scenarios and highlight its potential benefits and implementation challenges. 
     
    \item Focusing on a realistic farming case study for RIS-aided WPBUSNs, we specify the transmission protocol and jointly design the phase shifts of RIS and the time allocation among BC, WET, and WIT with the aim of maximizing the network throughput while considering the communication reliability characterized by a demodulated signal-to-noise ratio (SNR) threshold.
    
    \item According to the designed implementation guidelines, extensive simulation results demonstrate that the proposed RIS-aided WPBUSNs significantly enhances the network throughput compared to the three benchmarks. Moreover, we reveal that the performance gains achieved by RIS are influenced by parameters such as burial depth, VWC, the number of UDs, and the number of RIS elements. Furthermore, we outline the research challenges and potential future research directions for RIS-aided WPBUSNs.
\end{enumerate}

The reminder of this article is organized as follows. We first overview the recent research progress of WPBUSNs and RIS in Section II. Next, we introduce a general architecture of RIS-aided WPBUSNs, and discuss the potential benefits and challenges of synergizing RIS and WPBUSNs in Section III. We then present a case study of RIS-aided WPBUSNs in Section IV. Finally, we summarize the research challenges and further research directions in Section V, before drawing the conclusions in Section IV.

\section{Technical Background and Recent Development}
This section reviews the technical basis and recent research advancements for WPBUSNs and RIS, respectively.

\subsection{Wireless-powered Backscatter Underground Sensor Networks (WPBUSNs)}
BC does not rely on active communication components; instead, it allows devices to use instantaneously harvested energy to backscatter information by reflecting and modulating incoming RF signals through antenna impedance tuning. As a result, BC has emerged as a promising communication solution for energy-limited devices. Based on the implementation methods, the BC system includes three configurations: (i) \textit{Monostatic}, where a tag modulates the RF signal generated by a transceiver and then backscatters it back to the same transceiver; (ii) \textit{Bistatic}, where the transmitter and receiver are spatially separated; and (iii) \textit{Ambient}, where the RF signal comes from the ambient RF sources, such as TV/cellular tower, Bluetooth and WiFi AP, rather than relying on a dedicated RF source as in the bistatic configuration~\cite{BCMag}. Compared with the monostatic and ambient configurations, bistatic BC utilizes a dedicated RF source, ensuring that devices receive reliable and stable RF signals without interference from ambient sources. Therefore, considering the network reliability requirements in IoUT scenarios, this study focuses on the bistatic configuration.

Multi-user WPBUSNs based on the bistatic BC configuration is first introduced in~\cite{LinBSWPUSNs}. In this setup, UDs can not only harvest energy from the PS for subsequent WIT, but also backscatter information to the AP. Herein, it is necessary to design the optimal time allocation for BC, WET, and WIT operations so as to maximize sum throughput and facilitate urgent information transmission. However, one primary research challenge of WPBUSNs is that, in harsh underground environments, the throughput gains achieved by WPBUSNs are significantly reduced and even eliminated due to substantial losses in the subterranean wireless channel and the blockage issues in IoUT applications. To ensure the link quality of BC, increasing the transmit power of the PS or deploying multiple PSs is often inefficient and costly. One viable solution is to deploy RIS to compensate for blocked communication links and enhance the desired signal at the AP.

\subsection{Reconfigurable intelligent surface (RIS)}
An RIS is a planar surface composed of an array of software-controlled passive elements, each is capable of independently adjusting the phase shift and amplitude of incoming signals. Hence, RIS can reflect the incident signals toward the desired directions of the receiver, creating a virtual line-of-sight (LOS) propagation link between the transmitter and the receiver. Generally, RIS structures are categorized into two types: antenna-array-based and metasurface-based, depending on the materials used for the RIS reflecting elements~\cite{RISBCProceed}. The antenna-based RIS comprises a 2-D array of numerous low-cost antennas, with each antenna serving as a reflecting element that receives and reflects incident signals based on the antenna scattering principle. In contrast, the metamaterial-based RIS uses a metasurface composed of periodic metamaterial units, where each unit directly reflects the incident signals directly without receiving them. Thanks to its thin and dense characteristics, the metasurface-based RIS is easier and more flexible to be installed in IoUT scenarios~\cite{RISMag, RISUnderground}.

In a typical transmission network aided by a metasurface-based RIS, the transmitter first derives the optimal reflection coefficients for the RIS based on the estimated channel state information (CSI) and then sends the result to the RIS controller via a dedicated feedback link. Compared to traditional active relays, RIS operates at a lower cost regarding hardware and power consumption, as it passively reflects incident signals without requiring complex signal processing components or RF chains. Additionally, RIS operates in full-duplex without self-interference or thermal noise. Thanks to its low-cost and power-efficient features, RIS has been extensively studied in different applications, such as wireless-powered IoT networks, BC, non-terrestrial networks (NTN). It has demonstrated significant advantages in addressing research challenges within these domains. For instance, the authors in~\cite{RISWETMag} explore the self-sustainability of RIS in wireless-powered IoT networks to reduce energy harvesting time and improve system throughput. The study in~\cite{RISBCMag} proposes RIS-assisted BC to support massive IoT scenarios and demonstrated the performance gains of RIS in terms of success probability, throughput, and energy efficiency. Furthermore, in~\cite{RISNTNMag}, RIS is deployed on an unmanned aerial vehicle (UAV) to overcome terrain restrictions and enhance the reliability of BC in 6G NTN IoT scenarios.

While the aforementioned works demonstrate how RIS enhances system performance across various use cases, it remains unclear how RIS can be integrated into WPBUSNs, whether it is feasible for improving the system throughput, and what factors affect its performance gains, when an underground communication channel component is introduced. To answer these questions, in the following sections, we will first conceptualize a RIS-aided WPBUSNs system, and then discuss the potential benefits and implementation challenges of incorporating RIS into WPBUSNs, followed by an evaluation of its feasibility through a farming case study.

\section{RIS-aided WPBUSNs: Potential Improvements and Implementation challenges}

\subsection{General Architecture of RIS-aided WPBUSNs}
\begin{figure}[!t]
\centering
\includegraphics[width=3.4in]{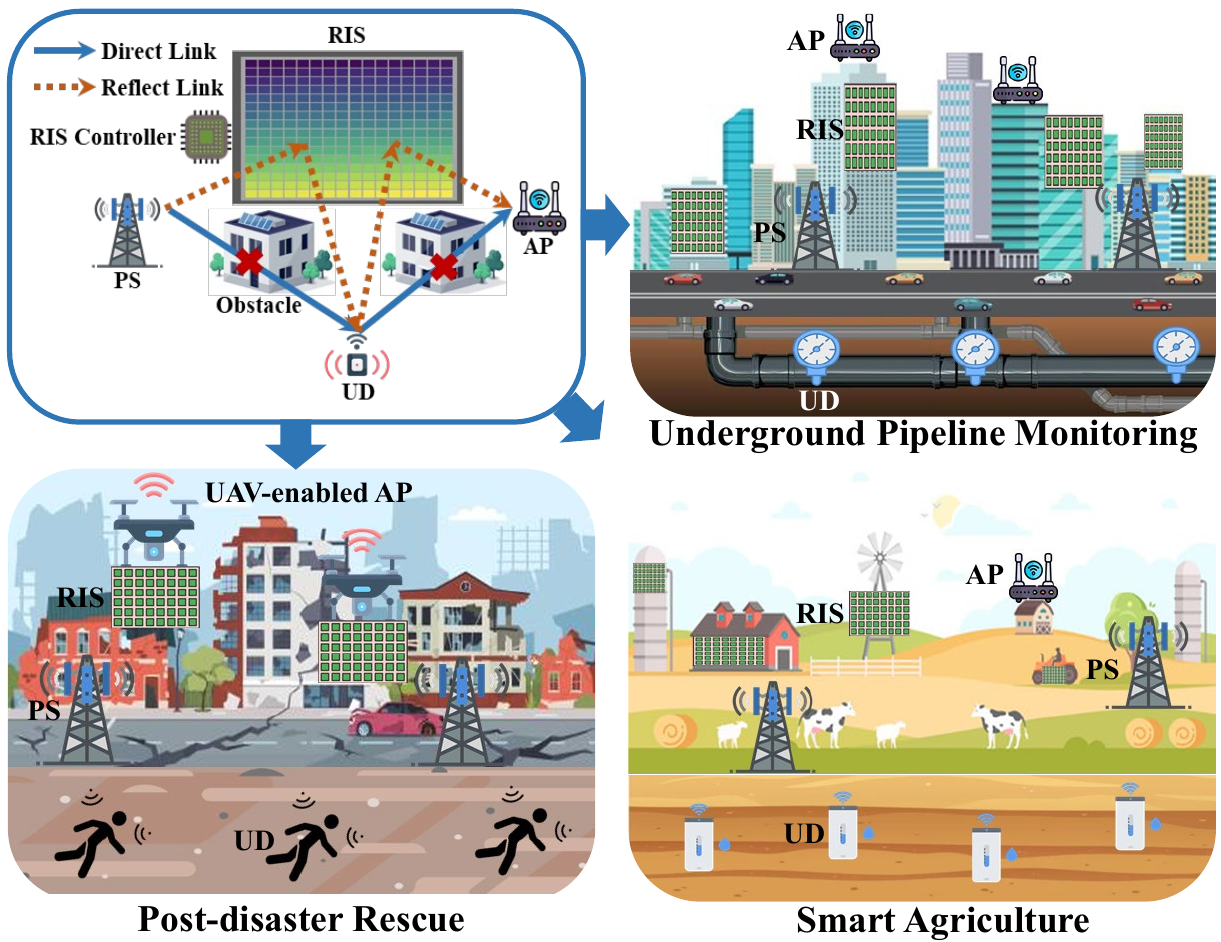}
\caption{A general architecture of our proposed RIS-aided WPBUSNs and its applications in underground pipeline monitoring, smart agriculture, and post-disaster rescue scenarios.}
\label{fig_generalarc}
\end{figure}

{Our proposed RIS-aided WPBUSNs architecture, as illustrated in~Fig.~\ref{fig_generalarc}, consists of PSs, RISs, energy-constrained UDs, APs, where each UD is equipped with an energy harvesting module, a BC circuit, and specific sensors tailored to meet the requirements of various underground monitoring scenarios. This allows the UDs to harvest energy from the PS for subsequent WIT in HTT mode, while backscattering the sensed data to the AP when the BC mode is activated. Notably, the UDs cannot operate in both HTT and BC modes simultaneously~\cite{LinBSWPUSNs}. Additionally, the AP is capable of demodulating the signals from both HTT and BC modes.} 

Considering the practical IoUT scenarios, the direct links between the PS and UDs, as well as the links between the UDs and AP, are obstructed by obstacles. Therefore, as shown in~Fig.~\ref{fig_generalarc}, RISs can be deployed on building walls, windmill supports, or UAVs, positioned between the PS and the AP. This configuration helps bypass these blockages, enhancing the channel gains for both uplink and downlink, and thereby improving the performance of WET, WIT, and BC. By further utilizing either orthogonal or non-orthogonal multiple access schemes, RIS-aided WPBUSNs can manage multi-UDs interference in the uplink and downlink, depending on the deployment scale and the specific requirements of the IoUT applications.

\subsection{Potential Benefits}
Three potential benefits that RISs can bring to WPBUSNs are outlined below:
\begin{itemize}
    \item \textbf{Enhanced Propagation Conditions:} The PS-to-UDs and UDs-to-AP links in IoUT are typically blocked by various obstacles, signals therefore undergo reflection, diffraction, and scattering, which degrades the quality of the received signal. By deploying RIS on flat surfaces (e.g., walls or outdoor signages) to reflect the signal around obstacles, we can then establish the strong LOS propagation conditions, thereby enhancing the channel conditions for WET, WIT, and BC. By such means, UDs can harvest sufficient energy from the PS, ensure reliable WIT, and increase throughput gain achieved by BC.
    
    \item \textbf{Network Coverage Extension:} Due to the harsh underground environment, WPBUSNs are restricted in large-scale underground monitoring scenarios. However, by installing RIS at the edge of the AP's coverage area, the communication range can be extended to reach UDs in previously unconnected regions. Consequently, RIS can significantly enhance the coverage of WPBUSNs.
    
    \item \textbf{Higher Energy Efficiency:} From an energy efficiency perspective, RIS can modify reflection coefficients to enhance the desired signal without the need for any active RF chains. Hence, the utilization of RIS is helpful for reducing the transmit power at the PS and effectively boosting the WET efficiency for UDs. Meanwhile, during the BC operation, UDs can harvest energy from both direct and RIS-reflected signals, allowing them to backscatter information while sustaining the circuit operations even in high-attenuation conditions. Additionally, the energy required for the CSI acquisition can be further reduced thanks to the enhanced channel gains of the PS-to-UDs and UDs-to-AP links achieved by RIS.    
\end{itemize}
\begin{figure}[!t]
\centering
\includegraphics[width=3.4in]{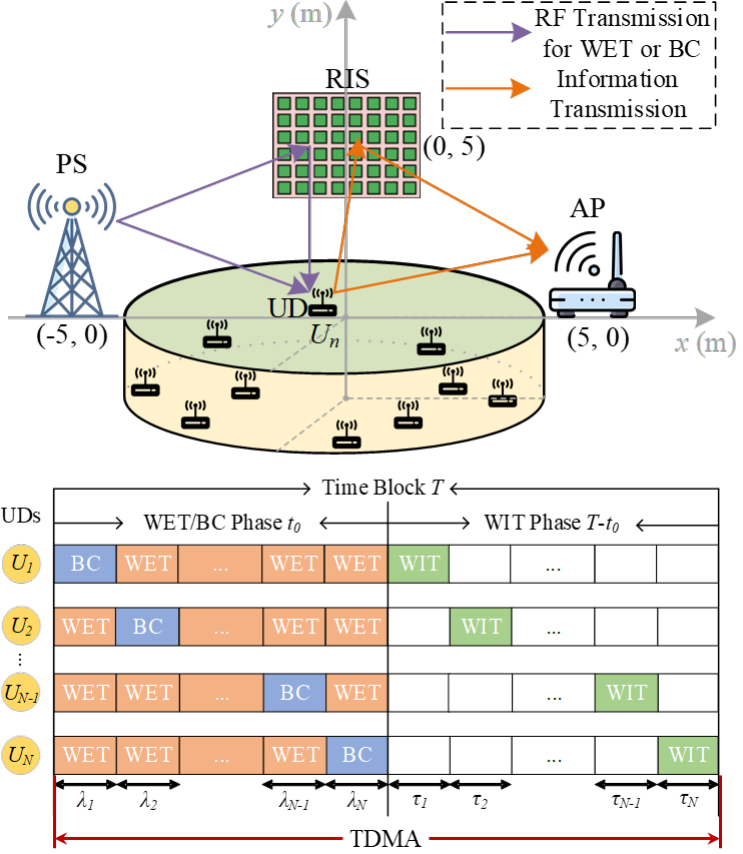}
\caption{A multi-user RIS-aided WPBUSNs system model considering the TDMA protocol.}
\label{fig_systemarch}
\end{figure}
\subsection{Implementation Challenges}
While RIS has great potential to enhance WPBUSNs, several concerns and challenges arise with its implementation in IoUT.

First, the channel model must be meticulously developed to accurately characterize the practical performance of the proposed RIS-aided WPBUSN system. Unlike terrestrial IoT, signal propagation in RIS-aided WPBUSNs occurs through soil, the soil-air interface, and air, and is significantly influenced by soil properties, burial depth, VWC, operating frequency, etc~\cite{Farmdata, LinLinkIOTJ}. Notably, attenuation at the soil-air interface varies with the propagation direction as soils have a higher permittivity than air. Moreover, due to different physical properties of soil, selecting an appropriate and accurate soil permittivity model is essential for precisely capturing underground path loss and effectively modeling RIS-aided communication channels. 

Second, the operation of WET, WIT, and BC, supported by RIS, should be carefully designed in the system. It is critical to design transceivers for RIS-aided WPBUSNs so as to avoid destructive wireless interference among WET, WIT, and BC operations. Although the authors in~\cite{LinBSWPUSNs} utilized the time division multiple access (TDMA) protocol to mitigate the interference, the impact of RIS on the protocol design has not yet been studied. The time allocation for the WET, WIT, and BC operations in the TDMA protocol needs to be jointly designed with the phase shifts of the RIS.

Third, a detailed implementation guideline should be developed to specify when and how to estimate the cascaded CSI, compute the optimal time allocation and phase shifts of RIS, and broadcast the solution to each UD. Furthermore, variations in soil's VWC due to irrigation and precipitation result in dynamic signal attenuation. This calls for the development of energy-saving mechanism that adaptively adjusts the transmit power of the PS and allocates system resources based on the current underground environment.

\section{A Farming Case Study for RIS-aided WPBUSNs}
To gain in-depth insight into the feasibility and performance of RIS in WPBUSNs, this section presents a case study on an RIS-aided WPBUSNs system considering a realistic center-pivot farm scenario, and provides the comprehensive analysis. 

\subsection{System Model and Simulation Setup}
A multi-user RIS-aided WPBUSNs system is depicted in Fig.~\ref{fig_systemarch}, comprising of a PS, $N$ UDs, an AP, and an RIS equipped with $K$ passive RIS reflecting elements. Herein, we consider that each UD, denoted by $U_n$ with $n=1, \ldots, N$, can adaptively switch between the HTT mode and the BC mode. The UDs are uniformly and randomly distributed within a circle of $5$-meter radius centered at $[0, 0]$, while the PS, RIS, and AP are located at the coordinates $[-5, 0]$, $[0, 5]$, $[5,0]$ m, respectively. Each of PS, UDs, and AP are equipped with a single antenna. We assume that the direct channels of PS-to-UDs and UDs-to-AP are blocked by obstacles, such as crops and farm machinery, while the other cascaded channels (i.e., PS-to-RIS, RIS-to-UDs, UDs-to-RIS, and RIS-to-AP) experience strong LOS propagation. Thus, the path loss exponents for the direct channels and cascaded channels are set to $3.2$ and $2$, respectively. Moreover, the LoRa modulation is utilized for decoding the signal from BC and HTT modes, thus the demodulated SNR threshold in our system is set as -20~dB~\cite{LinBSWPUSNs}. The key system parameters are listed in Table.~\ref{tab1}.

\begin{table}[t!]
\caption{Simulation Parameters and Configurations} 
\centering
\centering
\begin{tabular}{m{0.3\textwidth}<{\raggedright} m{0.15\textwidth}<{\centering}}%
\toprule
\textbf{Parameters}                & \textbf{Values}                    \\ \hline
\multicolumn{2}{l}{\textbf{Operation Environments~\cite{Farmdata, LinBSWPUSNs}}} \\ \hline
Total time block ($T$)               &1~s\\
Deployment radius ($R$)            & 5~m                                \\    
Number of UDs ($N$)	           &\textit{var} (64 by default)        \\
Height of PS and AP ($H$)            &1.5~m\\

Burial depth ($d_{u}$)             &\textit{var} (0.4~m by default)    \\
VWC ($m_v$)                                &\textit{var} (0.1 by default)       \\
Clay ($m_c$)                         &38\%     \\ \hline
\multicolumn{2}{l}{\textbf{Radio Configuration~\cite{Farmdata, LinBSWPUSNs}}} \\ \hline
Carrier center frequency ($f_c$)     &433~MHz                             \\
Bandwidth ($B$)                    &125~KHz                             \\
Antenna gains of PS, UDs and AP    &0~dBi            \\
Transmit power of PS ($P_{P}$) &30~dBm                              \\
Backscatter coefficient ($\eta_{n}$) &0.6                               \\
Energy conversion efficiency ($\varphi_{n}$)    &0.6                    \\
Noise power ($\sigma_{n}$)   &-117~dBm                            \\
Demodulated SNR threshold ($\gamma$) &-20~dB                      \\
Number of RIS's reflecting elements ($K$)      &\textit{var} (25 by default)\\
Path loss exponents of PS-to-RIS, RIS-to-UDs, UDs-to-RIS, and RIS-to-AP links         &2\\
Path loss exponents of PS-to-UDs and UDs-to-AP links&3.2\\
\bottomrule
\label{tab1}
\end{tabular}
\end{table}

As shown in Fig.~\ref{fig_systemarch}, TDMA, a classic orthogonal multiple access technique, is adopted in our RIS-aided WPBUSNs to avoid interference among UDs. During the WET/BC period $t_0$, the $n^{th}$ UD backscatters the signal emitted from the PS to the AP within its allocated time duration $\lambda_{n}$, while all other UDs simultaneously harvest energy from the PS. In the subsequent WIT period $T-t_0$, UDs utilize the harvested energy for the sequential WIT. Herein, the RIS is deployed to enhance the channel condition during the BC, WET, and WIT processes. The sum throughput is the total number of information bits transmitted by all UDs in both HTT and BC modes. To maximize the sum throughput, we derive the optimal time allocation and phase shifts of the RIS jointly for each UD in both HTT and BC modes based on the assumption of perfect CSI estimation, taking into account the saturation non-linear energy harvesting model~\cite{EHModel}, time constraints, and network reliability characterized by the demodulated SNR threshold at the AP. Meanwhile, to characterize the practical performance of our proposed RIS-aided WPBUSNs, we utilize the channel model introduced in~\cite{LinLinkIOTJ} so as to account for the accurate underground path loss model based on the mineralogy-based soil dielectric model, the refraction loss at the soil-air interface, and the cascaded channel introduced by the RIS. 

\begin{figure}[!t]
\centering
\includegraphics[width=3.4in]{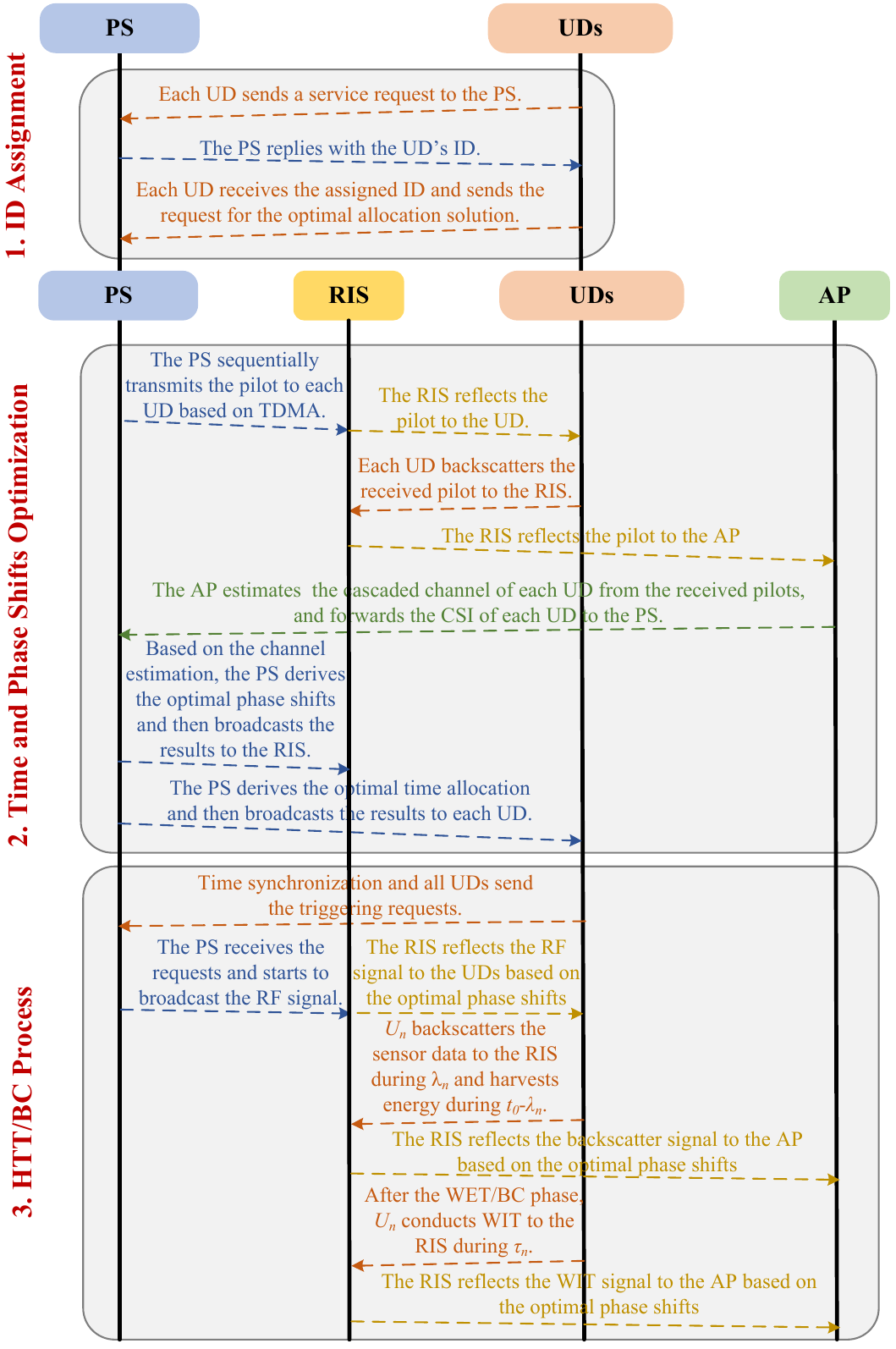}
\caption{The implementation diagram of our proposed RIS-aided WPBUSNs system.}
\label{fig_systemimple}
\end{figure}

Fig.~\ref{fig_systemimple} illustrates the practical implementation process for RIS-aided WPBUSNs with the optimal time allocation and phase shifts of the RIS. First, the PS assigns an identity number to each UD. Next, to estimate the cascaded channel of each UD, the PS sequentially transmits pilots to the corresponding UDs, which are then backscattered to the AP. The PS subsequently computes the optimal time allocation after determining the optimal phase shifts of the RIS based on the estimated CSI. Finally, the PS broadcasts the results of the optimal time duration and phase shifts to each UD and the RIS controller, allowing each UD to initiate the BC, WET, and WIT operations following time synchronization.

\subsection{Selected Numerical Results}
We compared the proposed RIS-aided WPBUSNs to the following benchmarks:
\begin{itemize}
    \item WPUSNs~\cite{LiuWPUSNs}: This is the traditional WPUSNs without RIS and BC. The time duration required for each UD's WET and WIT processes is carefully scheduled.
    
    \item WPBUSNs~\cite{LinBSWPUSNs}: The WPBUSNs system employs the optimal time allocation among WET, WIT, and BC,  but without the aid of RIS. 
    
    \item Random phase shifts~\cite{RISMag}: This is a RIS-aided WPBUSN and the phase shifts of the RIS are generated randomly within the range of $[0, 2\pi]$. The optimal time allocation among WET, WIT, and BC is considered in this benchmark.
\end{itemize}

\begin{figure}[!t]
\centering
\includegraphics[width=3.4in]{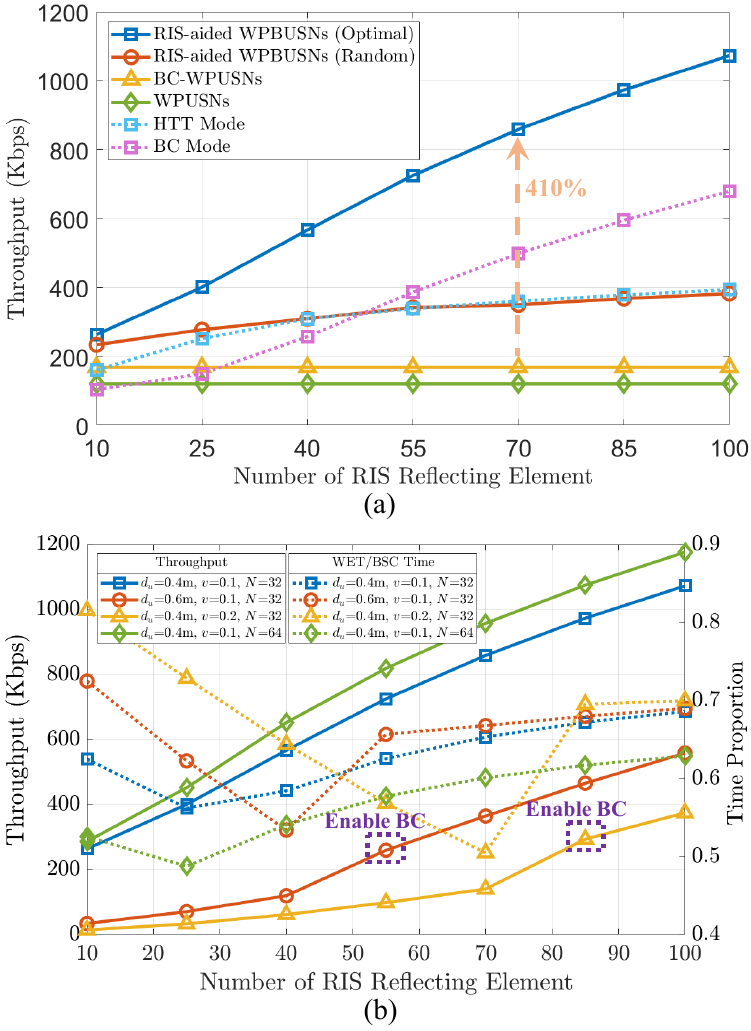}
\caption{Fig. 5. The selected numerical results. (a) The network throughput of RIS-aided WPBUSNs compared to the three benchmarks, as a function of the number of RIS reflecting elements $K$, where the dotted lines denotes the throughput produced by the HTT and BC modes, respectively, and (b) The network throughput and time allocation of our RIS-aided WPBUSNs as a function of the number of RIS reflecting elements $K$ under various burial depths, VWC and UDs’ number.}
\label{fig_res}
\end{figure}

As depicted in Fig.~\ref{fig_res}(a), the sum throughput of RIS-aided WPBUSNs with either optimal or random phase shifts increases with the number of RIS reflecting elements, whereas the overall throughput for WPUSNs and BC-WPUSNs remains approximately $120$ Kbps and $168$ Kbps, respectively. This is because an increased number of RIS reflecting elements enhances the channel gains of the PS-to-UDs and UDs-to-AP links, thereby boosting the capabilities of BC, WET and WIT. Furthermore, the higher performance gain of the proposed optimal phase shift approach is established with a larger number of reflecting elements, compared to the random phase shift approach. This underscores the fact that phase shift optimization can further enhance the received signal power and improve throughput. The throughput of the BC mode surpasses that of the HTT mode once the number of RIS reflecting elements exceeds $55$. This observation reveals that the the BC mode achieves a higher throughput gain compared to the HTT mode when there are more reflecting elements. Notably, our proposed RIS-aided WPBUSNs solution outperforms the three benchmarks in terms of the sum throughput, where RIS-aided WPBUSNs with the optimal phase shifts using $70$ RIS reflecting elements can achieve a $410$\% throughput increase compared to conventional WPBUSNs without RIS.

Dealing with smart agriculture applications, UNs need to buried at various depths to monitor the roots of different crops, while the VWC of soil varies with precipitation and irrigation. Hence, this study focuses on the effects of these two underground parameters on the system performance. Fig.~\ref{fig_res}(b) illustrates the effects of burial depths, VWC, and UDs' number on the sum throughput and time allocation of our proposed RIS-aided WPBUSNs system considering varying numbers of RIS reflecting elements. Herein,  From Fig.~\ref{fig_res}(b), we obtain the following three key observations.

First, the sum throughput decreases as the burial depth increases from $0.4$~m to $0.6$~m since a larger propagation path through underground soil leads to heightened attenuation in both uplink and downlink channels. When the number of RIS reflecting elements exceeds $55$, the proposed RIS-aided WPBUSNs enable BC at the burial depth of $0.6$~m; otherwise the system switches to the HTT-only strategy (i.e., a RIS-aided WPUSN without integrating BC) due to the unsatisfied BC reliability assurance. One can see that the time duration allocated for the WET/BC phase decreases with up to $40$ RIS reflecting elements, and then increases as the number of RIS reflecting elements surpasses $55$. The increased number of RIS reflecting elements enhances the channel conditions, resulting in reduced WET duration for the HTT-only strategy. On the other hand, the time assigned to WET/BC phase increases when the BC mode is enabled due to the higher throughput gains achieved by the BC mode. 

Second, the sum throughput deteriorates with VWC as a higher VWC results in larger attenuation constants of soil, leading to higher underground attenuation that significantly impacts the performance of WET, BC, and WIT. The proposed RIS-aided WPBUSNs requires over $85$ RIS reflecting elements to enable reliable BC at the VWC of $0.2$; otherwise the system defaults to the HTT-only strategy. From the time allocation results, it can be observed that the UDs invest less time to harvest energy in the HTT-only strategy as the number of RIS reflecting elements increases from $10$ to $70$. Moreover, since the BC is enabled with over $85$ RIS reflecting elements, more time is allocated to the WET/BC phase so as to achieve higher throughput gains.

Third, the sum throughput increases as more UDs are deployed to make full use of the RF resources from the PS in the WET phase. While UDs can harvest energy simultaneously, they cannot BC at the same time; thus, the system tends to allocate more time to the HTT mode as the number of UDs increases from $32$ to $64$.

\section{Challenges and Research Opportunities}
The above quantitative analysis demonstrates the feasibility of RIS in improving the sum throughput of WPBUSNs and enabling BC in the challenging underground scenarios, such as high-depth, high-VWC, and high-NLOS conditions. However, several potential challenges must be carefully addressed when using RIS-aided WPBUSNs in practice.
\begin{itemize}
    \item \textbf{Channel Configuration.} Multi-antenna PS and beyond-diagonal RIS (BD-RIS)~\cite{BDRISMag} can be utilized to deal with the high attenuation in underground soils. Herein, the digital beamforming at the PS, the scattering matrix of BD-RIS, and time scheduling should be carefully designed based on the estimated CSI to fully achieve the benefits of these techniques. Such optimization problem often faces highly non-convex optimization challenges. Consequently, compared to the traditional model-based solution or the optimization theory, leveraging reinforcement learning-based approaches enables adaptive resource allocation, thereby maximizing system throughput while ensuring network reliability in dynamic underground environments.
    
    \item \textbf{CSI Acquisition.} The estimated CSI is required at the PS for optimizing time allocation and phase shifts in RIS-aided WPBUSNs. However, due to the severe attenuation introduced by the underground soils and the inherent passive nature of RIS, reliable and perfect CSI acquisition becomes extremely challenging in massive underground monitoring scenarios. Therefore, it is essential to develop low-complex and efficient protocols for channel estimation to reduce the number of training pilots and minimize energy consumption among UDs. Additionally, the accuracy of CSI could be further enhanced through employing deep learning-based CSI acquisition approaches at the PS.
    
    \item \textbf{Interference Management.} Although TDMA can be used to avoid interference in BC and WIT processes, the latency and energy consumption required for time synchronization become prohibitively expensive when deploying massive UDs in RIS-aided WPBUSNs. Recently, non-orthogonal multiple access is identified as a promising strategy to support massive connectivity in the uplink. Herein, power allocation schemes and interference management strategies shall be carefully designed to ensure user fairness and throughput enhancement. Thanks to its powerful interference management capability, rate-splitting multiple access is worth exploring in the downlink~\cite{rsma}. Furthermore, fluid antenna multiple access emerges as a simpler, more scalable multiple access scheme capable of massive connectivity. 
    
    \item \textbf{Distributed RISs for Large-scale IoUT.} Only one fixed RIS may not be sufficient to provide reliable HTT and BC operations for large-scale IoUT scenarios. Therefore, one may consider deploying multiple RISs or mounting RISs on NTN platforms (e.g., UAVs) to enhance the signal quality for UDs without LOS connectivity and to cover the whole monitoring area. This inspires research on location optimization for RISs, effective trajectory planning of RIS-equipped UAVs, and optimal RIS scheduling for enhancing BC, WET, and WIT capabilities in RIS-aided WPBUSNs.  
\end{itemize}

\section{Conclusion}
To extend the recently emerged RIS-aided BC to the subterranean domain, this article proposed RIS-aided WPBUSNs for sustainable IoUT. After introducing the general architecture, potential improvements and implementation challenges for RIS-aided WPBUSNs, we assessed its feasibility and performance in a realistic farming scenario. Our numerical results demonstrated the proposed RIS-aided WPBUSNs realizes higher throughput than the three benchmarks. This was accomplished by optimally scheduling the time allocation and phase shifts of RIS while ensuring the network reliability. Our results also indicated that the network throughput and time allocation are significantly affected by underground parameters (i.e., burial depths and VWC), the number of UDs, and the number of RIS reflecting elements. Finally, we summarized the research challenges of RIS-aided WPBUSNs in terms of CSI acquisition, channel configuration, interference management, and distributed RISs, to motivate and guide further in-depth analytic and experimental investigations on this research domain.
\bibliographystyle{IEEEtran} 
\bibliography{ref}

\end{document}